\documentclass[conference,10pt]{IEEEtran}
\IEEEoverridecommandlockouts
\usepackage{cite}
\usepackage{amsmath,amssymb,amsfonts,bm}
\usepackage{algorithmic}
\usepackage{graphicx}
\usepackage{textcomp}
\usepackage{xcolor}
\usepackage{caption}
\usepackage{tikzscale}
\usepackage{siunitx}
\usepackage{tikz}
\usepackage{circuitikz}
\usepackage{pgfplots}
\usepackage{lipsum}  
\usepackage[caption=false]{subfig}
\usepackage[ruled,norelsize]{algorithm2e}
\usepackage[nolist,nohyperlinks]{acronym}
\usetikzlibrary{pgfplots.groupplots,shadows,shadows.blur,pgfplots.groupplots,calc,shapes.geometric,shapes.misc,arrows,matrix,fit,decorations.markings,shapes,positioning,patterns,shapes.arrows,backgrounds}
\usepgfplotslibrary{colorbrewer}
\pgfplotsset{compat=newest}
\DeclareUnicodeCharacter{2212}{−}

\def\BibTeX{{\rm B\kern-.05em{\sc i\kern-.025em b}\kern-.08em
    T\kern-.1667em\lower.7ex\hbox{E}\kern-.125emX}}
\begin{document}

\title{Instantaneous Bandwidth Estimation from Level-Crossing Samples via LSTM-based Encoder-Decoder Architecture
  \thanks{This work is supported by the German Ministry of Education and Research (BMBF) under the grants of 16KISK226 (6G-Health) and 16KISK016 (Open6GHub).}
}

\author{\IEEEauthorblockN{Johannes Königs, Carsten Bockelmann, Armin Dekorsy}
  \IEEEauthorblockA{Dept. of Communications Engineering \\
    University of Bremen, Bremen, Germany\\
    \{koenigs, bockelmann, dekorsy\}@ant.uni-bremen.de}
}

\maketitle

\begin{abstract}
  This paper presents an approach for instantaneous bandwidth estimation from \ac{LC} samples using a \ac{LSTM} encoder-decoder architecture.
  \ac{LC} sampling is a nonuniform sampling technique that is particularly useful for energy-efficient acquisition of signals with sparse spectra.
  Especially in combination with fully analog wireless sensor nodes, \ac{LC} sampling offers a viable alternative to traditional sampling methods.
  However, due to the nonuniform distribution of samples, reconstructing the original signal is a challenging task.
  One promising reconstruction approach is time-warping, where the local signal spectrum is taken into account. 
  However, this requires an accurate estimate of the instantaneous bandwidth of the signal.
  In this paper, we show that applying a \acl{NN} to the problem of estimating instantaneous bandwidth from \ac{LC} samples can improve the overall reconstruction accuracy.
  We conduct a comprehensive numerical analysis of the proposed approach and compare it to an intensity-based bandwidth estimation method from literature.  
\end{abstract}

\section{Introduction}
Low-power wireless sensor nodes have a wide range of potential applications, from environmental and industrial monitoring to patient monitoring. 
As nodes become increasingly energy-efficient, the range of possible applications is expected to expand further \cite{kandris2020applications}.
A key approach to power reduction in these resource-constrained applications is the use of efficient signal acquisition techniques.
One such method is \textit{event-based} sampling, where the signal is sampled only when an event occurs, so that more samples are taken at times of interest and fewer otherwise.
This is particularly useful for sparse-spectrum signals such as speech, electrocardiograms, and FM signals, where traditional fixed-frequency sampling techniques are suboptimal.
By effectively performing on-sensor analog compression, which decreases the required amount of data to be transmitted, the operating time of wireless sensor nodes can be extended.
Furthermore, event-based samplers such as the \acf{LC} sampler are realizable with a very low number of components.
However, signal reconstruction is challenging and requires advanced methods due to the nonuniform distribution of sampling times.
Methods include spline-based filtering \cite{greitans2007speech}, projections onto convex sets \cite{rzepka2018POCS} and a technique that combines minimum energy reconstruction with a method called \textit{time-warping} \cite{rzepka2018reconstruction}.

In this paper, we focus on the reconstruction from \ac{LC} samples using time-warping and minimum energy reconstruction.
We consider it promising due to its capability to accommodate for variations in local signal spectra which, as we will see later, can be linked to the level crossings of the signal.
The quality of this reconstruction depends heavily on an accurate estimate of the time-varying signal spectrum, or more specifically, an estimate of the instantaneous bandwidth $B(t)$.
The current state of research shows that there is a relationship between the intensity of level crossings and $B(t)$ \cite{rzepka2017estimation}.
Nonetheless, as of now, there exists no closed-form mathematical solution for the relationship between \ac{LC} samples and $B(t)$.
To address this challenge, we propose a method for processing nonuniform sequences of \ac{LC} samples in \acfp{NN} and apply a \acf{LSTM} encoder-decoder network to estimate $B(t)$.
This improves the overall signal reconstruction, as we will show numerically.

Notation: Since we are working with different types of signal samples, we denote the discrete uniform samples of the continuous signal $x(t)$ as $x(mT)$, where $m = 1 \ldots M$.
Furthermore, we define its nonuniform samples as the pair $\left\{t_n, x(t_n)\right\}$ with $n = 1 \ldots N$ and its \ac{LC} samples specifically as $\left\{t_k, x(t_k)\right\}$ with $k = 1 \ldots K$.

\section{Sampling Methods}
To illustrate the problem of signal reconstruction from \ac{LC} samples, we will first introduce the Nyquist-Shannon sampling theorem and then consider irregularly spaced samples.
We will then briefly discuss the minimum energy reconstruction, followed by the time-warping method of Clark et al. \cite{1164714}.
Finally, we consider a special case of nonuniform sampling: \ac{LC} sampling which is the focus of this work.

\subsection{Nyquist-Shannon Sampling Theorem}
Consider a bandlimited signal $x(t)$ that is bandlimited to the bandwidth $B$
\begin{equation}
    \forall |f| > B: \quad X(f) = \int_{-\infty}^{\infty} x(t) e^{-j2\pi f t}dt = 0.
\end{equation}
The Nyquist-Shannon sampling theorem states that a signal can be fully recovered from its samples $x(mT)$ taken at a constant rate of
\begin{equation}
    f_s = \frac{1}{T} = 2B
\end{equation}
by using the Whittaker-Shannon interpolation formula \cite{shannon2001mathematical}:
\begin{equation}
    x(t) = \sum_{m=-\infty}^{\infty} x(mT) \cdot \text{sinc}(2Bt - m).
    \label{eq:sh_interpolation}
\end{equation}

\subsection{Minimum Energy Reconstruction}
If we now sample $x(t)$ irregularly, resulting in the nonuniform samples $x(t_n)$, then (\ref{eq:sh_interpolation}) will not provide an error free reconstruction.

Instead, we can use minimum energy reconstruction \cite{yen1956nonuniform}.
We reconstruct $x(t)$ with a finite number of reconstruction functions
\begin{equation}
    x_M(t) = \sum_{m=1}^{M}\bm{c}[m] g_m(t)
\end{equation}
such that we minimize the error:
\begin{equation}
    \min_{\bm{c}} \lVert x(t) - x_M(t) \rVert^2 = e(\bm{c}).
\end{equation}
$\bm{c}$ can be found from the observations $x(t_n)$ by solving:
\begin{equation}
    x(t_n) = \sum_{m=1}^{M}\bm{c}[m] \text{sinc}{\left(2 B \left(t_n - t_m\right)\right)}
\end{equation}
for all $x(t_n)$.
This results in a linear equation system for all observations $N$:
\begin{equation}
    \bm{x} = \bm{G} \bm{c},
    \label{eq:eqsystem}
\end{equation}
where
\begin{equation}
    \bm{G}[n,m] = \text{sinc}{\left(2B(t_n-t_m)\right)} 
    \label{eq:recon_G}
\end{equation}
is an $N \times M$ matrix and $\bm{x}[n] = x(t_n)$ with $\bm{x}\in \mathbb{R}^{N\times1}$.
The solution is optimal in the \ac{MMSE} sense when the samples of $x(t)$ are corrupted by white noise \cite{choi2000stochastic}.

The matrix $\bm{G}$ is ill-conditioned \cite{choi2000stochastic}, especially when $\bm{G}$ is large or the sample times $t_n$ are highly nonuniformly distributed.
Thus, inversion can lead to large reconstruction errors, and it becomes necessary to introduce a regularization term \cite{rzepka2018reconstruction}:
\begin{equation}
    \left( \bm{G}^T \bm{G} + \varepsilon \bm{I}\right)^{-1} \bm{G}^T \bm{x} = \bm{c}.
    \label{eq:regularized}
\end{equation}

\subsection{Time-warping}
\label{sec:time_warping}
Clark et al. \cite{1164714} introduced an alternative method for reconstruction from nonuniform samples called time-warping.
They extend the Nyquist-Shannon sampling theorem by introducing a reversible transformation of the time axis.

First, we define a signal $x(\tau)$ that is bandlimited to $B = \frac{1}{2}$ in a bandwidth-normalized \textit{$\tau$-domain}.
The $\tau$-domain is characterized by a so-called time-warping function $\tau = \gamma(t)$, which is strictly monotonically increasing.
In the $\tau$-domain, we can describe the signal $x(\tau)$ by uniformly spaced samples $x(n)$, which with respect to $t$ are not necessarily uniform.
For a signal $y(t) = x(\gamma(t))$ and a linear $\gamma(t)$, the samples of $y(t)$ are uniformly distributed. 
If $\gamma(t) $ is not a linear function, they are nonuniformly distributed over $t$.

Substituting the time-warping function $\gamma(t)$ into (\ref{eq:sh_interpolation}) gives
\begin{equation}
    y(t) = x(\gamma(t)) = \sum_{n=-\infty}^{\infty} x(n) \cdot \text{sinc}(\gamma(t) - n).
\end{equation}
Since $\gamma(t)$ is a bijective function, we can represent the $\tau$-domain samples $x(n)$ as the samples $y(\gamma^{-1}(n))$ in the $t$-domain and obtain the time-warped interpolation formula:
\begin{equation}
    y(t) = \sum_{n=-\infty}^{\infty} y(\gamma^{-1}(n)) \cdot \text{sinc}(\gamma(t) - n).
    \label{eq:time-warping-interpolation}
\end{equation}
\begin{figure}
    \includegraphics[width=\linewidth]{plots/time_warping/plot.tikz}
    \caption{The signal $x(\tau)$ with its uniform samples $x(n)$ in the $\tau$-domain and its representation $y(t)$ in the $t$-domain with nonuniform samples $y(t_n)$.}
    \label{fig:time_warping}
\end{figure}
With (\ref{eq:time-warping-interpolation}), the signal $y(t)$ can be represented entirely by nonuniform samples $y(t_n)$.
The sample times are given by the integer crossings of $\gamma(t)$:
\begin{equation}
    t_n = \gamma^{-1}(n).
    \label{eq:sample_times}
\end{equation}
Accordingly, the derivative of $\gamma(t)$ describes the instantaneous sampling rate \cite{1164714}:
\begin{equation}
    f_s(t) = \frac{\delta \gamma(t)}{\delta t}.
\end{equation}
As in classical sampling, we assume that the sampling frequency is twice the bandwidth, but here both the sampling frequency and the bandwidth are functions of time $t$:
\begin{equation}
    B(t) := \frac{1}{2} \cdot \frac{\delta \gamma(t)}{\delta t}.
\end{equation}
Furthermore, if $B(t)$ is known, we can find $\gamma(t)$ by integration:
\begin{equation}
    \gamma (t) = 2 \cdot \int B(t) dt.
    \label{eq:intIband}
\end{equation}

Samples $t_n$ taken according to (\ref{eq:sample_times}) are more densely distributed when the instantaneous bandwidth $B(t)$ of $y(t)$ is high and are less dense when $B(t)$ is low.

In summary, time-warping allows reconstruction of nonuniformly sampled signals if $B(t)$ is known and the nonuniform samples are taken at the exact times we obtain from $\gamma(t)$.
In practice, neither $B(t)$ nor $\gamma(t)$ is known prior to signal acquisition, and therefore sampling from $\gamma(t)$ is not feasible.
Instead, we will consider the much simpler \ac{LC} sampler, which samples relative to the instantaneous bandwidth of a signal.

\subsection{Level-crossing Sampling}
A \acf{LC} sampler with a single level $L$ samples a signal $x(t)$ whenever the signal amplitude crosses the level $L$, yielding the sample $x(t_k) = L$.
The number of crossings in the interval $\left[0,T\right]$
\begin{equation}
    K_T(L) = \# \left\{t \in  \left[0,T\right] : x(t_k)=L \right\}
    \label{eq:nrcrossings}
\end{equation}
can be described by the Rice formula \cite{lindgren2013stationary}:
\begin{equation}
    \mathbb{E}\left[K_T(L)\right] = \frac{T}{\pi} \sqrt{\frac{\lambda_2^x}{\lambda_0^x}}\cdot e^{\frac{-L^2}{2\lambda_0^x}},
\end{equation}
where $\lambda_0^x = \text{var}\left[x(t)\right]$ and $\lambda_2^x = \text{var}\left[x^{(1)}(t)\right]$ are the first and second spectral moments of $x(t)$ and $x^{(1)}(t)$ is its first derivative.
Since the mean process bandwidth $\bar{B}$ is related to the spectral moments of $x(t)$
\begin{equation}
    \bar{B} = \sqrt{\frac{\lambda_2^x}{\lambda_0^x}}
\end{equation}
we have a direct link between the process bandwidth and the expected number of crossings.

Rzepka et al. showed in \cite{rzepka2017estimation} that given a nonstationary process $y(t)$  modeled as a stationary time-warped process $y(t) = x(\gamma(t))$, a similar relationship between the instantaneous bandwidth $B(t)$ and the expected number of \ac{LC} samples can be established:
\begin{equation}
    \mathbb{E}\left[K_T(L)\right] = \frac{1}{\pi} e^{\frac{-L^2}{2\lambda_0^y}} \int_0^T \sqrt{\frac{\lambda_2^y}{\lambda_0^y}} 2 \pi B(t) dt.
    \label{eq:mean_to_lbw}
\end{equation}

Although there is not yet a direct mathematical relation between the realization of a \ac{LC} sampling process and the instantaneous bandwidth $B(t)$, (\ref{eq:mean_to_lbw}) suggests that such a connection exists.

\section{Reconstruction from level-crossing samples}
\ac{LC} samples are highly irregularly distributed.
They are taken with respect to $B(t)$, but not exactly according to $\gamma(t)$, so neither the minimum energy nor the time-warping reconstruction leads to a satisfactory reconstruction.

However, \cite{rzepka2018reconstruction} presents a combined method that first transforms the \ac{LC} samples from the $t$-domain to the $\tau$-domain and then applies the minimum energy reconstruction.
To this end, we modify (\ref{eq:recon_G}) to incorporate the time-warping function, resulting in the time-warped sinc matrix
\begin{equation}
    \bm{G}[k,n] = \text{sinc}{\left(\gamma(t_k)-n \right)}
\end{equation}
for $K$ \ac{LC} samples and $N$ nonuniform samples derived from the time-warping function $\gamma(t)$.
Then again we need to solve (\ref{eq:eqsystem}) for $\bm{c}$.

The method is thus an attempt to transform an irregular sampling sequence into a regular one.
Warping the samples back to the $\tau$-domain may not necessarily result in a uniform grid because the \ac{LC} samples are not directly derived from the time-warping function $\gamma(t)$.
However, this process can aid in the minimum energy reconstruction by increasing the uniformity of the sample spacing.
In practice, $\gamma(t)$ may not be perfectly known and needs to be derived from an estimation of $B(t)$.
Since no closed form mathematical solution exists, we believe that an estimation using a \acf{NN} is a promising alternative to kernel based estimators such as \cite{rzepka2017estimation}.

\subsection{Instantaneous Bandwidth Estimation using an LSTM Encoder-Decoder}
We formulate the problem of estimating $B(t)$ as finding a function $f(\cdot)$ that produces an estimate $\hat{B}(t)$ of $B(t)$ in the interval $t \in \left[0,T\right]$, given a sequence of \ac{LC} times and amplitudes $\left\{t_k, x(t_k)\right\}$.
Suppose $B(t)$ is a bandlimited function so that it can be described by $M_B$ samples $B(mT_B)$ taken at a rate of $f_{s,B} = 1/T_B$.

We can express the function with its trainable parameters as:
\begin{equation}
    \hat{\bm{b}} = f(\bm{s}, \bm{\Theta}),
\end{equation}
where $\hat{\bm{b}}\in \mathbb{R}^{M_B}$ are the $M_B$ samples describing $\hat{B}(t)$ and $\bm{s}\in \mathbb{R}^{K \times 2}$ is the sequence pair of crossing times and amplitudes in vector form.
To find $f(\bm{s}, \bm{\Theta})$, we apply supervised learning so that the weights are learned such that
\begin{equation}
    \min_{\bm{\Theta}} \lVert \bm{b} - \hat{\bm{b}} \rVert^2.
\end{equation}

To minimize the estimation error, we want to apply sequence-to-sequence learning with an \ac{LSTM} encoder-decoder architecture.
This architecture was first described in \cite{sutskever2014sequence} for natural language processing tasks.
We chose this architecture for its ability to handle temporal relationships and sequences of different lengths.
This is necessary since, depending on the realization of $x(t)$ and the level configuration, the length of the sample sequences $\bm{s}$ vary drastically in a fixed time interval.
To train the network in batches of multiple realizations of $\bm{s}$, we pad them to a fixed length $P$.

The network consists of a masking layer to mask the padded values in $\bm{s}$, a bidirectional \ac{LSTM} layer acting as the encoder, a repeat vector layer, an \ac{LSTM} layer as the decoder, and a final time-distributed dense layer.
The bidirectional encoder \ac{LSTM} layer encodes the sequence into a single feature vector that is repeated by the next layer for each time step of our output sequence $\bm{b}$.
It is then fed into the decoder \ac{LSTM} layer along with the hidden states $\bm{h}$ and the cell states $\bm{c}$ of the encoder.
A final dense layer is then applied to each time step output of the \ac{LSTM}, resulting in a fixed size sequence of length $M_B$.

For the \ac{LSTM} layers, we used the tanh function as the activation, the sigmoid function as the recurrent activation and, a linear activation function for the time-distributed dense layer.
To train the network, we used Adam optimization \cite{kingma2017adammethodstochasticoptimization} with a learning rate of 0.001 and a gradient clipping value of 1.
We trained and tested the network with different hyperparameters, varying the units and layers of the encoder and decoder.
For the final network we chose a single bidirectional \ac{LSTM} encoder layer with 256 units and a single bidirectional decoder layer with 512 units.

\section{Numerical Simulation}
Now, we will analyze the reconstruction performance of the time-warped minimum energy reconstruction method using our proposed network for estimation of $B(t)$.
Since we use supervised learning, we need a test signal with known instantaneous bandwidth $B(t)$.

After training the network, we want to compare its performance with the reconstruction using the estimation method proposed in \cite{rzepka2017bandwidth} and with a simple piecewise linear interpolation reconstruction.
We chose to use the \ac{NMSE} as our error measure to compare $J$ signals $x_j(t)$:
\begin{equation}
    \text{NMSE} \left\{x\right\} = \frac{1}{J \cdot M} \sum_{m=1}^{M} \sum_{j=1}^{J} \frac{\left| x_j(mT) - \hat{x}_j(mT) \right|^2}{\left| x_j (mT) \right|^2}.
\end{equation}

\subsection{Test Signals}
\label{sec:testsig}
\begin{figure}
    \includegraphics[width=\linewidth]{plots/bandwidth_examples/plot.tikz}
    \caption{Realizations of instantaneous bandwidths $B(t)$ with random amplitudes for different $\Upsilon$.}
\end{figure}

\begin{figure}
    \includegraphics[width=\linewidth]{plots/LCS_to_DoF_N_L/plot.tikz}
    \caption{Total number of crossings over the number of nonuniform samples $t_n$ required to describe a test signal $x(t)$ in the interval $\left[0,T\right]$.}
    \label{fig:LCS_to_DoF_N_L}
\end{figure}
We generate our test signals with known instantaneous bandwidths by applying the time-warping framework.
We will first generate $B(t)$ from random samples, then using the time-warping function $\gamma(t)$ find the nonuniform samples $t_n$ and randomly select amplitudes for the samples. With (\ref{eq:time-warping-interpolation}) we then find one realization of a test signal $x(t)$.

For our test signals, we assume that they are bounded and that the instantaneous bandwidth functions $B(t)$ are bandlimited to $\Upsilon$ and fully defined on the time interval $t \in [0, \SI{1}{\second}]$.
Accordingly, we choose $M_B=2.1 \cdot \Upsilon + 1$ samples $B_m$ for a slight oversampling, allowing us to define $B(t)$:
\begin{equation}
    B(t) = \sum^{M_B}_{m=1} B_m \text{sinc}(2\Upsilon t-m).
\end{equation}
We choose random $B_m \in \mathcal{U}\left(\SI{5}{\hertz}, \SI{100}{\hertz}\right)$ from the uniform distribution and select only $B_m$ that result in a positive $B(t)$.

We can determine the number of samples $N$ that describe a single test signal $x(t)$ from the time-warping function $\gamma(t)$:
\begin{equation}
    N = \int \gamma(t) dt.
\end{equation}
Then, we select $N$ samples $x_n \in \mathcal{N}\left(0, 1\right)$ from the normal distribution such that $-4 \leq x(t) \leq 4$ and finally use (\ref{eq:time-warping-interpolation}) to construct the signal.

We want to observe the influence of the complexity of $B(t)$ and the sampler configuration on the estimation and reconstruction.
For $B(t)$ we choose $\Upsilon \in \left[1,2,3,4,5,10,20\right] \si{\hertz}$ to keep the number of instantaneous bandwidth samples $M_B$ and thus the network size small.
For the \ac{LC} sampler, we vary the number of levels $N_L \in \left[5,6,7,8,9,10,13\right]$ and place them evenly between the signal boundaries $[-4, 4]$.
We assume that the resulting values of the \ac{LC} samples $t_k$ are continuous, hence no quantization in time takes place.
For each pair of parameters we generate 200000 realizations, resulting in \ac{LC} sequences which we pad to a fix length $P=512$.
We then split the realizations into a training (70\%), validation (20\%) and test dataset (10\%).

Fig. \ref{fig:LCS_to_DoF_N_L} shows the average of the total number of level crossings
\begin{equation}
    \bar{K} = \frac{1}{U}\sum_{u=0}^{U} \sum_{i=0}^{N_L} K_{T,u}(L_i)
\end{equation}
for all $U$ test signal realizations $x_u(t)$ that are constructed from the same number of samples $N$ for the different sampler configurations.
According to the time-warping method, in the best case we only need $N$ samples taken at the exact times $t_n$ derived from the time-warping function $\gamma(t)$ to perfectly reconstruct our signal.
From Fig. \ref{fig:LCS_to_DoF_N_L} we can see that $N_L=7$ is the point where our sampler theoretically produces enough values to perfectly reconstruct our signal.
Since a \ac{LC} sampler only samples relative to $B(t)$, it is improbable that samples will be taken at times $t_n$, making a perfect reconstruction highly unlikely.
For $N_L>7$ the samplers produce more samples, and for $N_L=13$ we produce twice as many samples as are minimally needed for reconstruction.
Since we are interested not only in the signal reconstruction error, but also in the estimation of $B(t)$ from the \ac{LC} samples, we decided to include the results for $N_L<7$.

\subsection{Comparison Algorithm}
We chose to compare our \ac{NN} estimator with the estimation algorithm presented by Rzepka et al. \cite{rzepka2017estimation}, which relies on local estimates of the \ac{LC} intensity.
This version improves on their previous estimator \cite{rzepka2014recovery}, addressing concerns regarding positive bandwidths and bandlimitedness.
However, as noted by the authors, it still has stability issues and problems with different level configurations.

The algorithm in \cite{rzepka2017estimation} has a number of parameters, that need to be selected.
For our numerical analysis, we set the clipping value of the intensity function $\varepsilon_0 = \SI{0.002}{\hertz}$ to prevent negativ bandwidth values from occuring and set the power ratio to $r=1/\sqrt{3}$ as described in \cite{rzepka2017estimation}.
It also requires knowledge of $\Upsilon$ for the projection onto bandlimited sets, which is provided accordingly.
We determined the regularization term of (\ref{eq:regularized}) for the minimum energy reconstruction numerically to be $\varepsilon = 0.05$ for both estimation approaches.

Furthermore, we want to compare our reconstruction performance with a basic piecewise linear interpolation between samples.
To interpolate in the interval $[t_0, t_1)$ between two samples $x(t_0)$ and $x(t_1)$, we use
\begin{equation}
    x(t) = x(t_0) + \frac{x(t_1) - x(t_0)}{t_1 - t_0} (t - t_0)
\end{equation}
and repeat this process for all \ac{LC} samples $x(t_k)$.

\section{Results}
In Fig. \ref{fig:estimation}, the estimation error of $B(t)$ for both methods is shown for different $N_L$ over the bandwidth $\Upsilon$ of $B(t)$.
Higher $\Upsilon$ values result in a more complex $B(t)$, while higher $N_L$ produce more samples for the estimation.
Both methods show an increase in estimation error as the complexity of $B(t)$ increases.
Notably, Rzepka's method exhibits superior performance for $N_L=5$ compared to sampler configurations with higher level count.
One possible explanation is the stability issue mentioned above when crossing higher value levels.
For the \ac{NN} estimator we can observe a small increase in estimation performance with higher number of levels $N_L$.

Fig. \ref{fig:reconstruction} shows the corresponding reconstruction error for selected $\Upsilon$.
Comparing the two methods and piecewise linear interpolation, the reconstructions improve as the number of levels $N_L$ increases.
The overall reconstruction error decreases when using the \ac{NN} estimator compared to Rzepka's estimator.
The reconstruction with Rzepka's estimator performs worse than a simple piecewise linear interpolation, indicating that the estimation quality is not sufficient to aid the minimum energy reconstruction.
This is also evident for $\Upsilon = \SI{20}{\hertz}$, where both estimation methods perform similarly in the reconstruction, suggesting that $B(t)$ is too complex to estimate from the given samples.
Any improvements are most likely due to the increase in samples from which to reconstruct.

In comparison, the \ac{NN} estimator reconstruction outperforms the piecewise linear interpolation method for most $N_L$, but is overtaken by it in some cases for $N_L=13$.
While both methods improve with a larger number of samples, the time-warping approach degrades the reconstruction result especially for higher $\Upsilon$, as is evident for $\Upsilon = \SI{20}{\hertz}$, where both $B(t)$ estimates result in a higher reconstruction error than piecewise linear interpolation.

Fig. \ref{fig:NMSE_all_N_L_X_selection} compares the performance between reconstruction with \ac{NN} estimation, full a priori knowledge over $B(t)$ and piecewise linear interpolation for $\Upsilon=\SI{1}{\hertz}$.
At $\Upsilon=\SI{1}{\hertz}$, the \ac{NN} estimator performs best, with $\text{NMSE} \left\{B\right\} = \SI{-19.6}{\decibel}$ for $N_L=10$.
Although the error of the instantaneous bandwidth estimation is quite small in this case, the reconstruction with a priori knowledge shows a significant performance improvement, suggesting that a much more accurate estimation is required for optimal reconstruction.

\begin{figure}
    \includegraphics[width=\linewidth]{plots/NMSE_all_W_avg_B/plot.tikz}
    \caption{Estimation error of the instantaneous bandwidth function $B(t)$ over varying complexity (higher bandwidth $\Upsilon$ of $B(t)$) and different numbers of levels.}
    \label{fig:estimation}
\end{figure}

\begin{figure}
    \includegraphics[width=\linewidth]{plots/NMSE_all_N_L_X_2024_11_05/plot.tikz}
    \caption{Signal reconstruction error over different number of levels for the \ac{NN} estimation method, for Rzepka and for piecewise linear interpolation.}
    \label{fig:reconstruction}
\end{figure}

\begin{figure}
    \includegraphics[width=\linewidth]{plots/NMSE_all_N_L_X_selection/plot.tikz}
    \caption{Signal reconstruction error for reconstructions with estimated, a priori, and no knowledge about $B(t)$ for $\Upsilon=\SI{1}{\hertz}$ over $N_L$.}
    \label{fig:NMSE_all_N_L_X_selection}
\end{figure}

\section{Conclusion and Outlook}
This paper presents an approach for instantaneous bandwidth estimation from \ac{LC} samples using an \ac{LSTM} encoder-decoder architecture for synthetic signals.
By increasing the accuracy of instantaneous bandwidth estimation, the paper demonstrates improved signal reconstruction.
The comparison with existing methods highlights possible performance improvements for the time-warped minimum energy reconstruction, but also shows that high errors occur when the instantaneous bandwidth cannot be estimated precisely.

We also showed that the combination of time warping and the minimum energy reconstruction performed better than a simple piecewise linear interpolation in certain scenarios.
We believe that further improvements are necessary to fully utilize this approach.
One possible improvement is to use the additional information provided by \ac{LC} sampling, such as time-dependent bounds for either the reconstructed signal or the samples $\bm{c}$, since both are bounded by the surrounding levels.
Furthermore, since the supervised learning approach presented only works for synthetic signals with known instantaneous bandwidths, this method is not applicable to practical signals such as electrocardiograms.
To this end, in future work, we want to investigate whether training a \ac{NN} that minimizes the signal reconstruction error calculated from the time-warped minimum energy reconstruction is a possible optimization goal.
This would allow us to apply the method to practical signals and may provide a better understanding of the relationship between the \ac{LC} samples and the instantaneous bandwidth.

\begin{acronym}[FDMA]
\acro{ADC}{analog-to-digital converter}
\acro{ASR}{average sampling rate}
\acro{ECG}{electrocardiogram}
\acro{MSE}{mean squared error}
\acro{NMSE}{normalized mean squared error}
\acro{STFT}{short-time Fourier transform}
\acro{EMD}{empirical mode decomposition}
\acro{VBW}{variable bandwidth}
\acro{EEG}{electroencephalography}
\acro{NN}{neural network}
\acro{UWB}{ultra-wideband}
\acro{LSTM}{long short-term memory}
\acro{MMSE}{minimum mean squared error}
\acro{LC}{level-crossing}
\end{acronym}

\bibliographystyle{IEEEbib}
\bibliography{include/references}

\end{document}